# Angular factor corrections in thin film x-ray diffraction

T. R. S. Prasanna

Department of Metallurgical Engineering and Materials Science

Indian Institute of Technology, Bombay

Powai, Mumbai – 400076, India

**Abstract**

The current practice of angular (Lorentz, Polarization and absorption (LPA)) corrections in bulk materials for higher angle broad peaks, due to the Kα doublet, are based on the results of Pike (Acta Cryst., 12 (1959) 87-92) and Ladell (Acta Cryst. 14 (1961) 47-53). These results are extended to thin film x-ray diffraction. The current practice in thin film studies of using the absorption factors for monochromatic radiation in LPA corrections is incorrect in both symmetric and grazing incidence x-ray diffraction. The correct expressions for absorption corrections are developed. In addition, a new length scale, the linear regime, is shown to exist for very thin films where the absorption factor and the integrated intensity are independent of the absorption coefficient. For very thin films, it is practically advantageous to use the expressions for the linear regime over those for the thin film regime. The procedures to obtain both the main results are of general validity and can be extended to several thin film diffraction techniques.

## 1.0 Introduction

X-ray diffraction techniques are the primary tools for the structural characterization of crystalline materials. The most common technique is the flat-plate x-ray diffraction of polycrystalline materials (Birkholz, 2005; Cullity, 1978; Guinebretiere, 2007; Pecharsky and Zavalij, 2009; Warren, 1990; Waseda *et al.*, 2011). Most thin film x-ray diffraction techniques are based on the flat-plate geometry (Birkholz, 2005; Fewster, 1996).

The fundamental difference between the integrated intensity of bulk and thin films is due to the different absorption factors. In thin films, the absorption factor is dependent on the angle, and the product of the linear absorption coefficient ($\mu$) and the thickness (t). The $\mu t$ product is the defining feature of thin film diffraction (Birkholz, 2005).

As is well known, in laboratory diffractometers, higher angle peaks are broadened due to the presence of the K$\alpha$ doublet. In accurate peak profile studies, including for accurate lattice parameter determination, several corrections must be made before analysis.

In many studies on bulk materials, the raw data is pre-treated before line profile analysis is performed (Delhez *et al.*, 1977; Fitzpatrick *et al.*, 2005; Iwashita *et al.*, 2004; Miyazaki *et al.*, 1992; Somers *et al.*, 1989; Tian and Atzmon, 1999; Tian and Zhang, 2009). They include, background corrections, Lorentz-Polarization (LP) and absorption (A) factor corrections followed by K$\alpha_2$ stripping.

For thin films as well a similar pre-treatment is performed (Birkholz, 2005; Liu et al. 2015; Guazonne, 2006; Skrzypek and Baczmański, 2001; Ely et al. 1999). Thus, for both, bulk materials and thin films, angular corrections are made to the higher angle diffraction data before peak profile analysis.

The theoretical basis of the angular factor corrections in bulk materials was developed by Pike and Ladell (Pike, 1959; Ladell, 1961; Pike and Ladell, 1961). Indeed, Ladell's (1961) result is the basis of the current practice for angular corrections in bulk materials (Delhez *et al.*, 1977).

In this article, we extend the main results of Pike and Ladell (Pike, 1959; Ladell, 1961; Pike and Ladell, 1961) developed for bulk materials to thin films. In particular, we present new analysis of the absorption factor for thin films. It leads to two important results. Firstly, the absorption factor currently used as part of the angular corrections in peak profile analysis is incorrect and the correct expression is presented. Secondly, for very thin films, the absorption factor can be represented by its linear approximation which is independent of the μt product. Hence, very thin films can be categorized to belong to a new category based no thickness, the linear regime. That is, depending on the thickness, materials response to x-rays falls in three categories, bulk regime, thin film regime and linear regime.

**2.0 Angular factors affecting the integrated intensity of thin films**

The expression for the integrated intensity for a homogenous material (with constant linear absorption coefficient) is derived in standard texts and is given by (Birkholz, 2005; Cullity, 1978; Guinebretiere, 2007; Pecharsky and Zavalij, 2009; Warren, 1990; Waseda *et al.*, 2011)

$$I_{hkl} = I_0 \left(\frac{\lambda_m^3 \, m_{hkl} \, |F_{hkl}(T)|^2}{V_{uc}^2}\right) \left(\frac{1 + cos^2 2\theta_{hkl}}{sin\theta_{hkl} \, sin2\theta_{hkl}}\right) \left(\frac{1 - e^{-\frac{2\mu t}{sin\theta_{hkl}}}}{2\,\mu}\right) \quad (1)$$

We have included several constants in $I_0$ for convenience. The other symbols have their usual meanings – $\lambda_m$ is the wavelength of the monochromatic radiation, $m_{hkl}$ is the multiplicity factor and $d_{hkl}$ the inter-planar spacing of planes with Miller indices (*hkl*), $F_{hkl}(T)$ is the temperature dependent structure factor, $\mu_{\lambda m}$ is the linear absorption coefficient for wavelength $\lambda_m$, *t* is the thickness, $V_{uc}$ is the volume of the unit cell and $\theta_{hkl}$ is the angle that satisfies the Bragg's law $\lambda = 2\, d_{hkl}\, sin\theta_{hkl}$. The linear absorption coefficient is a product of the mass absorption or attenuation coefficient ($\mu_m$) and the density, $\mu = \mu_m\, \rho$.

For bulk materials, the thickness is sufficiently large so that µt » 1 and the absorption factor (last term in Eq.1) reduces to A = 1/2µ. While, in principle, it is exact for $t \to \infty$, in practice this condition is satisfied for thickness beyond 10 - 100 microns, depending on the material. In all other cases, the material is considered to be in the thin film regime. Therefore, *the µt product is the defining feature of thin film x-ray diffraction* (Birkholz, 2005).

The trigonometric terms in Eq.1 are due to different factors (Birkholz, 2005; Cullity, 1978; Guinebretiere, 2007; Warren, 1990; Waseda *et al.*, 2011) and are collectively referred to as the Lorentz-Polarization (LP) factor. It is represented below separately

$$LP_{hkl}^{mono} = \frac{1 + cos^2 2\theta_{hkl}}{sin^2\theta_{hkl}\, cos\theta_{hkl}} \qquad (2)$$

The factor of 2 in sin2θ in Eq.1 is kept out since it does not contribute to angle dependence.

The above expressions, Eq.1 and Eq.2, are valid for unpolarized x-rays when no monochromator is used. The polarization term will change if a monochromator is used (Birkholz, 2005; Pecharsky and Zavalij, 2009). However, since in this work we extend the results of Pike (1959) and Ladell (1961) to thin films, we use the same expressions above that were derived by them for the case with no monochromator. The main results are unaffected by use of a monochromator and can be easily re-derived for this experimental condition.

Eq.1 is for <u>strictly monochromatic</u> radiation (Pike, 1959; Ladell, 1961). In reality, the emission lines have a spectral profile. Frequently, the average valued of the wavelength in the Kα doublet is used along with mass absorption coefficient for this average wavelength. For example, the International Tables of Crystallograpy (Creagh and Hubbell, 2006) reports the mass absorption coefficients of various elements for Cu K$\bar{\alpha}$ radiation i.e. $\mu_m(\bar{\alpha})$, along with its energy, which translates to the average wavelength, $\lambda(\bar{\alpha})$. It is clear that they are constant values averaged over the Cu Kα doublet.

However, in laboratory diffractometers, Kα$_1$ and Kα$_2$ radiations are present and both have a spectral profile. As is well known, the spectral profile leads to broad peaks at higher angles. The works of Pike and Ladell (Pike, 1959; Ladell, 1961; Pike and Ladell, 1961) discuss the angular factor corrections necessary to determine accurate peak positions (and lattice parameters) from broad higher angle peaks.

Pike (1959) showed that for the case of a spectral profile e.g. Kα doublet, the $\lambda^3$ term in Eq.1 (that is no longer constant) must be substituted with Bragg's law. Combining this with the angular dependence in Eq.1, the LP factor in this case is given by Eq.16 of Pike (1959)

$$LP_{hkl}^{\lambda} = \left((tan\theta)\left(1 + cos^2 2\theta_{hkl}\right)\right) \tag{3}$$

Pike's (1959) work implicitly assumed a constant absorption coefficient by not considering its wavelength dependence. Given that Eq.16 of Pike was in the context of the Kα doublet, the mass absorption coefficient reported in the International Tables for K$\bar{\alpha}$ radiation, i.e. $\mu_m(\bar{\alpha})$, for the absorption factor can be combined with Eq.3 above (Eq.16 of Pike)'s result. Further, it can be generalized to thin films by using $\mu_m(\bar{\alpha})$ in the thin film absorption factor. This leads to the angular factor generalized for thin films to be

$$LPA\,(\lambda, \bar{\alpha}) = LP(\lambda)\,A(\bar{\alpha}) = \left((tan\theta_{hkl})\left(1 + cos^2 2\theta_{hkl}\right)\right)\left(\frac{1 - e^{-\frac{2\mu_m(\bar{\alpha})\rho t}{sin\theta_{hkl}}}}{2\,\mu_m(\bar{\alpha})\,\rho}\right) \tag{4}$$

Subsequently, Ladell (1961) pointed out that that the mass absorption coefficient, $\mu_m$, is not constant but varies as $\lambda^3$ and cancels the $\lambda^3$ term in Eq.1. Thus, the angular factor is the same as in Eq.1 in this case (Ladell, 1961; Pike and Ladell, 1961). That is, the angular factor including the Lorentz, Polarization and Absorption (LPA factor) is given by

$$LPA^{bulk}(\lambda) = \frac{1 + \cos^2 2\theta_{hkl}}{\sin^2 \theta_{hkl} \cos \theta_{hkl}} \tag{5}$$

It is readily seen that $LPA^{bulk}(\lambda)$, Eq.5 is the same as the $LP_{hkl}^{mono}$, Eq.2. We note that $LPA^{bulk}(\lambda)$, Eq.5, is currently used for angular corrections in <u>bulk materials</u> (Delhez *et al.*, 1977; Fitzpatrick *et al.*, 2005; Iwashita *et al.*, 2004; Miyazaki *et al.*, 1992; Somers *et al.*, 1989; Tian and Atzmon, 1999; Tian and Zhang, 2009). It is most important to note that it actually also includes the absorption correction, even though mathematically is the same as $LP_{hkl}^{mono}$. This is because $LP_{hkl}^{mono} = LPA^{bulk}(\lambda)$ or Eq.2 = Eq.5.

This important result of Ladell (1961) is also applicable to thin film x-ray diffraction and its extension is discussed below. We note that in the case of thin film diffraction there are two sources of the $\lambda^3$ dependence of the absorption coefficient in Eq.1. Following Ladell (1961) the mass absorption coefficient can be represented as $\mu_m = c_\mu \lambda^3$. For the $1/2\mu$ term in Eq.1, this substitution cancels the $\lambda^3$ dependence of $I_{hkl}$, as shown by Ladell (1961) for bulk materials.

However, in the case of thin films, this substitution must also be made in the exponential term in Eq.1. Since, in this case the wavelength, $\lambda$, (in the K$\alpha$ doublet) is variable and un-cancelled, following Pike (1959) the $\lambda^3$ term in $\mu_m = c_\mu \lambda^3$ must be substituted by Bragg's Law, $\lambda = 2\, d_{hkl} \sin\theta$. With these substitutions, the absorption factor generalized to thin films is given by

$$A_{dispersion}^{tf}(\theta) = \frac{(1 - e^{-16\, c_\mu\, \rho\, d_{hkl}^3\, \sin^2\theta\, t})}{2\, c_\mu\, \rho} \tag{6}$$

In Eq.6, following Ladell (1961) the $\lambda^3$ dependence of the $1/2\mu$ term cancels the $\lambda^3$ term in Eq.1 and is hence, dropped from the denominator.

The LPA factor is given by

$$LPA^{thin\ film}(\lambda) = \left(\frac{1+\cos^2 2\theta_{hkl}}{\sin^2\theta_{hkl}\cos\theta_{hkl}}\right)\frac{(1-e^{-16\,c_\mu\,\rho\,d^3_{hkl}\sin^2\theta\,t})}{2\,c_\mu\,\rho} \qquad (7)$$

We now consider the current practice of LP(θ) A(θ) corrections for <u>thin films</u>. Birkholz (2005) describes (p-97, p-250) the current practice of LPA corrections for <u>thin films</u> as "*it can thus be concluded that for broader peaks the correction of the measured intensity by division by LP(2θ) A(2θ) should be performed prior to fitting of reflections.*" The current practice of angular corrections for peak profile analysis includes the Lorentz-Polarization (LP(2θ)) and absorption (A(2θ)) terms and are given by (Birkholz (2005; Liu et al. 2015; Guazonne, 2006; Skrzypek and Baczmański, (2001); Ely et al. 1999)

$$LPA^{current} = \left(\frac{1+\cos^2 2\theta_{hkl}}{\sin^2\theta_{hkl}\cos\theta_{hkl}}\right)\left(1-e^{-\frac{2\mu_m(\bar{\alpha})\rho t}{\sin\theta_{hkl}}}\right) \qquad (8)$$

It is readily seen that Eq.8 follows directly from Eq.1 that is for <u>monochromatic radiation.</u>

A comparison Eq.8, the currently used LPA factor, with Eq.4 and Eq.7 shows that it is different from both the LPA factors obtained by extending the results of Pike (Eq.4) and Ladell (Eq.7) to <u>thin films</u>. The presence of the Kα doublet causes peak broadening. If a constant absorption coefficient is used, then the LP factor obtained by Pike must be used (Pike 1959; Pike and Ladell 1961). That is, the LPA factor given by Eq.4 that is obtained by extending Pike's result to thin films must be used. If the wavelength dependence of the absorption coefficient is incorporated, then the LPA factor obtained by extending Ladell's result, Eq.7 must be used. In neither case can the expression for monochromatic radiation, Eq.8, be used to correct broad peaks that are due to the presence of the Kα doublet. Thus, it is clear that *there is no theoretical justification for the use of Eq.8 to make LPA corrections for thin film diffraction data*. Thus, it is reasonable to conclude that the currently used expression for LPA corrections, Eq.8, is incorrect since it differs from both Eq.4 and Eq.7.

We note that the angular factor used in current practice, Eq.8, is correct for integrated intensity calculations. *Thus, for thin films, the angular factors to be used for integrated intensity applications (Eq.8) and peak broadening corrections (Eq.7) are different*. In contrast, for bulk materials, the angular factors for integrated intensity calculations (Eq.2) are the same as for peak broadening corrections (Eq.5) since Eq.2 = Eq.5. This is an important distinction between LPA corrections for bulk and thin films that must be recognized.

**3.0 Uncertainties in the absorption factor (for spectral profile) and its implications**

The LPA factor to be used for angular corrections of broad peaks in peak profile analysis is given by Eq.7. In particular, the absorption factor is given by Eq.6. There are several parameters that need to be determined to obtain the absorption factor, Eq.6. There are uncertainties in their determination.

The constant $c_\mu$ can be easily estimated from the relation $\mu_m = c_\mu \lambda^3$. It can be determined from the data given in the International Tables of Crystallograpy (Creagh and Hubbell, 2006) using the relation $\mu_m(\bar{\alpha}) = c_\mu \lambda^3(\bar{\alpha})$. For example, for Si using Cu K$\bar{\alpha}$ radiation, $\mu_m(\bar{\alpha}) = 63.7$ and $\lambda(\bar{\alpha}) = 1.5418$ Å leading to $c_\mu = 17.38$.

It is well known (Birkholz, 2005; Fewster, 1996; Pecharsky and Zavalij, 2009) that there are uncertainties in determining the density, ρ, and the thickness, t. These uncertainties are also present in the determination of the absorption factor for current practice, i.e. for constant wavelength (Birkholz, 2005; Fewster, 1996; Pecharsky and Zavalij, 2009). These uncertainties are present in both the correct expression, Eq.7, and the present expression, Eq.8.

However, a new and important source of uncertainty in the correct expressions, Eq.6 and Eq.7, that is absent in Eq.8, is that the interplanar spacing, $d_{hkl}$, must also be known. We recall that the angular corrections are necessary to obtain correct peak positions from which accurate lattice parameters and interplanar spacings are determined. Thus, it is clear that the value of $d_{hkl}$ that can be used in Eq.6 is approximate. Since the absorption factor varies as $d_{hkl}^3$ uncertainties in $d_{hkl}$ are magnified in Eq.6.

It is also clear that there is circularity in the use of $d_{hkl}$. Using an approximate $d_{hkl}$ as part of LPA corrections cannot directly lead to an accurate $d_{hkl}$ or lattice parameter from the corrected peak profiles.

This naturally suggests an iterative approach for LPA corrections to obtain corrected peak profiles. That is, the $d_{hkl}$ obtained from the corrected peak profiles must be substituted in the expression for the LPA and the corrected peak profiles recalculated. This process must be reiterated until the $d_{hkl}$ values used in the LPA corrections and those obtained from the corrected peak profiles are in agreement to the desired accuracy. This is clearly a cumbersome process.

**4.0 Linear approximation of the absorption factor – the linear regime**

From the above discussion, it is clear that the presence of the $d_{khl}$ term in the absorption factor makes the LPA corrections cumbersome. For thicker films, there is no alternative but to adopt the iterative procedure. However, for thinner films an alternative approach is possible that eliminates the determination of the uncertainties of several parameters discussed above in making LPA corrections.

In the limit $t \to 0$, the exponential term in Eq.6 can be approximated by Taylor expansion upto the linear term, $\exp(-x) \approx 1- x$ or $1-\exp(-x) \approx x$. This leads to the absorption factor $8\, d_{hkl}^3 \sin^2\theta\, t$. The constant factor ($8\, d_{hkl}^3\, t$) has no angle dependence and after combining all trigonometric terms the angular factor in Eq.7 is modified in the linear regime and is given by

$$LPA^{lin\,reg}(\lambda) = \left(\frac{1 + cos^2 2\theta_{hkl}}{cos\theta_{hkl}}\right) \qquad (9)$$

It is immediately evident that $LPA^{lin\,reg}(\lambda)$ is independent of the density, $\rho$, the thickness, t and also the interplanar spacing, $d_{hkl}$. Its advantages are discussed later below. It is also important to note that Eq.9 is independent of $c_\mu$, i.e. $LPA^{lin\,reg}(\lambda)$ is independent of the absorption coefficient. This is unlike for the thin film regime where $LPA^{thin\,film}(\lambda)$ depends on the absorption coefficient.

**4.1 Thickness range of validity of the linear regime**

The thickness range of validity of the linear regime needs to be determined. We note that in $LPA^{thin\,film}(\lambda)$ is due to the wavelength distribution is due to the presence of the Kα doublet. Since the absorption coefficient varies with wavelength, the thickness range also varies. This suggests that it is much simpler to estimate the thickness range of validity using the maximum wavelength present in the Cu Kα doublet ~ 1.546 Å from the expression for the absorption factor in Eq.1, $(1-\exp(-2\mu t/\sin\theta))/2\mu$. Thus, it leads to a conservative estimate of the thickness range of validity of the linear regime since $\mu_m(\lambda=1.546\,\text{Å})$ is greater than the mass absorption coefficient for all wavelengths present in the Cu Kα doublet.

A further simplification is possible. Instead of calculating $\mu_m(\lambda=1.546\,\text{Å})$, we can calculate the thickness for 1.01 $\mu_m(\bar{\alpha})$ for Cu Kα doublet, since this is greater than $\mu_m(\lambda=1.546\,\text{Å})$. The advantage is that this $\mu_m(\bar{\alpha})$ is tabulated in International Tables. For example, for Si $\mu(\lambda = 1.546) = 64.22$ and 1.01 $\mu_m(\bar{\alpha}) = 64.34$.

In summary, even though there is a range of wavelengths present, the thickness range of validity of the linear regime can be estimated from the absorption factor for constant wavelength by using a constant mass absorption coefficient that is 1.01 $\mu_m(\bar{\alpha})$.

The linear approximation, exp(-x) = 1-x, leads to ≤ 1% error for x ≤ 0.135. Using this criterion, $\tau_{lr}$, the thickness for which the linear regime is valid is given by $\tau_{lr}$ = 0.135 sinθ/2μ. Fig.1 plots the absorption factor, (1-exp(-2μt/sinθ)), and its linear approximation, 2μt/sinθ, for ZnO(100) peak with μ = 1.01 $\mu_m(\bar{\alpha})$. It shows that the 1% error criterion is a reasonable approximation.

For the lowest angle peaks in MgO(111), Si(111), ZnO(100), SrTiO₃(100), FePt(002) and FePt(001), the linear regimes extend to $\tau_{lr}$ ~ 2100 nm, 1100 nm, 680 nm, 240 nm, 88 nm and 44 nm respectively. For higher angle peaks, $\tau_{lr}$ increases because of the increase in sinθ. This is for Cu-K$_\alpha$ (λ=1.5418 Å) radiation which is the most common choice for thin film diffraction (Birkholz, 2005). *Clearly, the linear regime is of practical importance.*

For the above calculations, the mass absorption coefficients, $\mu_m(\bar{\alpha})$, were taken from International Tables for Crystallography (Creagh and Hubbell, 2006). The angle and density data were taken from ICDD Card Nos. 45-0946 (MgO), 27-1412 (Si), 36-1451 (ZnO), 35-0754 (SrTiO₃) and 43-1359 (FePt).

This criterion is an arbitrary choice based on acceptable error in the linear approximation. If the criterion is set to be ≤ 0.5% error, the thickness is given by $\tau_{lr,0.5}$ = 0.10 sinθ/2μ. In this case, the linear approximation is valid for thicknesses below 1500 nm - 30 nm for the above examples. For ≤ 0.1% error, the thickness is given by $\tau_{lr,0.1}$ = 0.045 sinθ/2μ. For this case, the linear approximation is valid below 700 nm – 15 nm for the above examples. Thus, even for 0.5% and 0.1% errors due to the linear approximation, the linear regime is of practical importance. The linear regime can be extended by using higher energy radiations for which mass absorption coefficients are lower (Creagh and Hubbell, 2006).

The above criterion is based on the assumption that the absorption factor, $(1-\exp(-2\mu t/\sin\theta))/2\mu$, is exactly known. In practice, it must be obtained from experiments and consequently, and there are associated errors in determining the relevant quantities (Birkholz, 2005; Fewster, 1996; Lhotka *et al.*, 2001; Pecharsky and Zavalij, 2009; Qiao *et al.*, 2004). Therefore, the linear approximation can be used at least until the errors in both approaches are similar.

**4.2 Advantages of the linear regime over the thin film regime for LPA corrections**

It is important to note that $LPA^{lin\,reg}(\lambda)$ (Eq.9) is derived from $LPA^{thin\,film}(\lambda)$ (Eq.7) for very small thicknesses and both are mathematically equivalent. Thus, for very thin films either expression can be used. However, from a practical standpoint $LPA^{lin\,reg}(\lambda)$ is much more advantageous as discussed below.

Firstly, it is unnecessary to estimate several parameters (density, $\rho$, thickness, t and the interplanar spacing, $d_{hkl}$) in order to make angular corrections using $LPA^{lin\,reg}(\lambda)$ unlike for $LPA^{thin\,film}(\lambda)$. Secondly, because the $d_{hkl}$ term is no longer part of the angular factor, it is no longer necessary to use an approximate $d_{hkl}$ to obtain the corrected peak profile. Thus, the iterative procedure suggested above is no longer necessary to obtain corrected peak profiles. For these reasons, it is clearly advantageous to work with the linear regime rather than the thin film regime for very thin films for peak profile corrections.

## 5.0 Integrated intensity in the linear regime

The above discussion was restricted to LPA corrections for peak profile analysis of thin films. However, the concept of the linear regime can be applied to the expression for the integrated intensity as well. This follows from Eq.1 the general expression for the integrated intensity of a thin film.

In the limit $t \to 0$, the exponential term in Eq.1 can be approximated by Taylor expansion upto the linear term, $\exp(-x) \approx 1 - x$. Thus, in the linear regime, the absorption factor becomes, $A_{\theta 2\theta} = t/\sin\theta$. Substituting $A_{\theta 2\theta} = t/\sin\theta$ in Eq.1 gives the integrated intensity in the linear regime as

$$I^{lr}_{hkl} = I_0 \left( \frac{\lambda_m^3 \, m_{hkl} \, |F_{hkl}(T)|^2}{V_{uc}^2} \right) \left( \frac{1 + \cos^2 2\theta_{hkl}}{\sin^2 \theta_{hkl} \, \sin 2\theta_{hkl}} \right) t \qquad (10)$$

Eq.10 implies that, for finite thickness $t$ of the thin film, the integrated intensity is obtained from the whole irradiated volume of the film without attenuation and with different angle dependence. *It is also independent of the μt product, the defining feature of thin film diffraction* (Birkholz, 2005).

In Eq.10, all angular factors are combined together as is the usual practice. This is called the LP factor for thin films in the linear regime and given by

$$LP^{lr}_{hkl} = \left( \frac{1 + \cos^2 2\theta_{hkl}}{\sin^3 \theta_{hkl} \, \cos\theta_{hkl}} \right) \qquad (11)$$

A comparison of Eq.1 and Eq.10 shows that the intensity ratios of peaks are different in bulk (A = 1/2μ) and in the linear regime because of the extra $1/sin\theta_{hkl}$ dependence in the latter. They are related by

$$\frac{I^{lr}_{h_1k_1l_1}}{I^{lr}_{h_2k_2l_2}} = \frac{I^{bulk}_{h_1k_1l_1}}{I^{bulk}_{h_2k_2l_2}} \frac{sin\theta_{h_2k_2l_2}}{sin\theta_{h_1k_1l_1}} \quad (12)$$

As a theoretical principle, the expressions derived above for the linear approximation, Eq.10 to Eq.12, are exact for $t \to 0$. The practical range of validity of the linear approximation for < 1% error is given by $\tau_{lr}$ = 0.135 sinθ/2μ as discussed above, except that in this case μ = $\mu_m(\bar{\alpha})$ and not μ = 1.01 $\mu_m(\bar{\alpha})$. Since they differ only by 1%, the same (lower) thickness calculated above can be used. Thus the range obtained earlier, 44 nm – 2000 nm, is valid for both cases.

We note that for higher angle peaks that are broadened due to the Kα doublet, the $\lambda^3$ term in Eq.10 must be substituted by Bragg's law (Pike, 1959; Ladell, 1961). This leads to an additional $sin^3\theta$ term for the angular factor that when combined with the angular factors already present in Eq.10 leads to Eq.9. That is, $(sin^3\theta) LP^{lr}_{hkl} = LPA^{lin\ reg}(\lambda)$ or Eq.9 can also be obtained from the Eq.10, the basic expression for the integrated intensity in the linear regime. Thus, Eq.10, Eq.11 and Eq.9 are the linear regime equivalent of Eq.1, Eq.2 and Eq.7 respectively where the latter are for thin film and bulk regimes.

**5.1 Advantages of the linear regime over thin film regime for integrated intensity applications**

As discussed earlier, for very thin films, both the expressions for the integrated intensity, Eq.1 and Eq.10 mathematically equivalent and valid. Indeed, Eq.10 has been obtained from Eq.1 with a linear approximation. The question of whether Eq.10 has practical advantages needs to be addressed. We have already discussed advantages of the linear regime over the thin film regime for peak profile applications.

For integrated intensity applications, using the expression for the thin film regime, Eq.1, requires determination of the density and thickness of the film. There are errors associated with their determination (Birkholz, 2005; Fewster, 1996; Lhotka *et al.*, 2001; Pecharsky and Zavalij, 2009; Qiao *et al.*, 2004). In contrast, it is unnecessary to determine these quantities when Eq.10, the integrated intensity in the linear regime is used. This is a clear advantage.

Another advantage of the using the expression for the integrated intensity in the linear regime, Eq.10, over the expression for the thin film regime, Eq.1, is in full profile analysis e.g. Rietveld fitting of the diffraction pattern.

In Rietveld fitting of diffraction patterns it is well known (McCusker et al. 1999; Pecharsky and Zavalij, 2009; Volz et al. 2006) that the angle dependent absorption factor is strongly correlated with the thermal displacement factor. This effect is seen when the Debye-Scherrer method is used to obtain diffraction data where the absorption factor is angle dependent. This effect is important in Rietveld refinement of both x-ray (McCusker et al. 1999) and neutron diffraction (Volz et al. 2006) data.

In the case of symmetric Bragg-Brentano geometry of bulk materials the absorption factor is not angle dependent and is constant and becomes part of the scale factor (Pecharsky and Zavalij, 2009). However, for thin films, the absorption factor is angle dependent as seen in Eq.1. Thus, the correlation of absorption factor with thermal displacement factor in the Rietveld fitting of thin film diffraction data is a concern (Pecharsky and Zavalij, 2009, p.193-196).

In contrast, if Eq.10, the expression for the integrated intensity in the linear regime, is used instead of Eq.1, for very thin films, it is seen that the problem of correlation of the absorption factor with the thermal displacement factor is eliminated because Eq.10 is independent of the absorption coefficient. Thus, Rietveld refinement in the linear regime becomes similar to that for bulk materials in that the angle dependent absorption factor is absent. This is clearly advantageous over using the expression for the thin film regime, Eq.1.

**6.0 The linear regime is the third regime along with bulk and thin film regimes**

The above discussion shows that the general expression for the integrated intensity, Eq.1, which is valid for all thicknesses, is altered at both extremes. The bulk regime is exact for $t \rightarrow \infty$ but in practice begins beyond t ≈ 10 – 100 microns depending on the material. Similarly at the other extreme, $t \rightarrow 0$, the linear approximation is exact.

In particular, at both extremes, the integrated intensity is independent of the μt product, the defining feature of thin film diffraction. As seen earlier, Eq.10, Eq.11 and Eq.9 are the linear regime equivalent of Eq.1, Eq.2 and Eq.7 respectively where the latter are for thin film and bulk ($t \rightarrow \infty$) regimes. Thus, similar to the separate categorization of bulk materials, it is justified to place thin films in the linear regime in a separate category.

The advantages of using the expressions for the linear regime over those of the thin film regime, especially for LPA corrections have already been discussed. The above discussion shows that in practice, similar to the case of bulk regime, the linear regime extends well beyond the theoretical limit and is valid for t ≈ 45 nm - 2000 nm depending on the material. It can also be increased by using higher energy radiations.

Table 1 summarizes the LPA factors in all three regimes, bulk, thin film and linear regime to be used for angular corrections in peak profile analysis and for integrated intensity applications. It is clear that the linear regime belongs in a separate category because of the very different dependences of the LPA factors compared to the other two categories.

**7.0 Analyses for bulk materials can be extended to thin films in the linear regime**

It is readily seen that in the linear regime, the integrated intensity and the LPA factor (for peak profile corrections) are both independent of the absorption coefficient and the thickness, i.e. the $\mu t$ product, unlike the expressions for the thin film regime. In this regard, the expressions are similar to those for the bulk regime. Thus, many of the concepts developed for analysis of bulk regime can be extended to the analysis in the linear regime.

Two of them have already been discussed. First, the LPA corrections in both the bulk and linear regimes are independent of the $\mu t$ product. Secondly, the Rietveld refinement in the bulk and linear regimes do not require the absorption coefficient or the thickness as a parameter. Another similarity is discussed below.

FePt thin films are important for ultra-high recording density applications (Maret *et al.*, 2012; Yang *et al.*, 2012). In current devices, the thicknesses are 60 nm or less (Yang *et al.*, 2012). The linear regime (1% error) is ~44 nm for (001) and ~87 nm for (002) peaks of FePt. Therefore, most thin films are likely to be in the linear regime.

In particular, the degree of ordering, determined by the order parameter, S, is very important for this application. The order parameter is obtained from the ratio of intensities of the (001) superlattice line to the (002) fundamental line, $I_{001}/I_{002}$, and is a function of thickness of the film. Yang *et al.* (2012) discuss this issue in detail. We compare the results obtained using the present method with their approach for randomly oriented samples.

Table II summarizes primary diffraction data obtained from Yang *et al.* (2012) and in addition, the results obtained from the present approach for the linear regime. For bulk materials, $\frac{I_{hkl}(001)}{I_{hkl}(002)} = 2.08$, from Eq.1 (as in Yang *et al.* (2012).

For thin films, the order parameter as a function of thickness is given by (Yang *et al.*, 2012)

$$S = \left(\frac{I_{001} \, I_{002}^{theo,bulk} \, (1 - e^{-0.0016t})}{I_{002} \, I_{001}^{theo,bulk} \, (1 - e^{-0.0032t})}\right)^{1/2} \tag{13}$$

where t is the thickness in nanometers. The thin film absorption factor is included as seen from the (1-exp-2μt/sinθ) term in Eq.13. For $t \to \infty$, Eq. 13 gives the bulk values of the order parameter (Yang *et al.* 2012).

Using the bulk theoretical ratio, 2.08, the order parameter can be obtained for the measured intensities of the (001) and (002) peaks for thin films as a function of thickness. For example, if the experimental ratio is $\frac{I_{001}}{I_{002}} = 2$ for a 40 nm thick film, the order parameter is obtained as S(40 nm) = 0.70 from Eq.13.

In the present approach, if the thin film is in the linear regime, the theoretical intensity ratio can be directly obtained using Eq.10. Table II shows that $\frac{I^{lr}_{001}}{I^{lr}_{002}} = 4.16$. The order parameter is given by

$$S^{lr} = \left(\frac{I_{001} I^{lr}_{002}}{I_{002} I^{lr}_{001}}\right)^{1/2} \tag{14}$$

Eq.14 is similar in structure to the expression for bulk order parameter, Eq.13, with $t \to \infty$. If the experimental ratio is $\frac{I_{001}}{I_{002}} = 2$, the order parameter in the linear regime is $S^{lr}$ = 0.69. This is very close to the value of 0.70 obtained using the full expression, Eq.13. Clearly, the linear approximation gives order parameter that is comparable to the full calculation. We note that the errors in the μt product that will influence the final result have not been considered in the calculation using the full expression, Eq.13.

The above examples show that analyses developed for bulk materials can be applied to thin films in the linear regime. This is because in both regimes, the expressions for the integrated intensity are independent of the μt product.

## 8.0 Extension to grazing incidence x-ray diffraction (GIXRD)

The main results above were obtained from the new analysis of the absorption factor. They are based on ideas that are of general validity and are applicable to other thin film diffraction techniques as well. The extension to grazing incidence x-ray diffraction (GIXRD) is discussed below.

In GIXRD, the integrated intensity has $\lambda^3$ dependence (Breiby et al. 2008). For very low angles of incidence that are near the critical angle, refraction corrections have to be made using standard expressions (Dummer et al. 2000; Liu et al. 2015, Wronski et al. 2009). The LP factor in GIXRD is given by (Dummer *et al*. 2000; Liu *et al*. 2015; Wronski *et al*. 2009)

$$LP_{\alpha 2\theta} = \left(\frac{1 + cos^2 2\theta}{sin^2 \theta}\right) \qquad (15)$$

The absorption factor in GIXRD is given by (Birkholz, 2005; Gloaguen et al. 2014; Lhotka *et al.*, 2001)

$$A_{\alpha 2\theta} = \left(\frac{1 - e^{-\mu t k_\alpha}}{\mu\, k_\alpha\, sin\alpha}\right) \qquad (16)$$

where α is the constant angle of incidence and $k_\alpha$ = 1/sinα + 1/sin(2θ-α).

The current practice of angular corrections for peak profiles in GIXRD uses Eq.16 for absorption correction (Birkholz, 2005; Liu et al. 2015; Skrzypek and Baczmański, 2001; Wronski et al. 2009).

To correct for peak broadening of higher angle peaks, the recipe is clear. Following Pike (1959) and Ladell (1961), the mass absorption coefficient must be represented by $\mu_m = c_\mu \lambda^3$. This substitution for the $\mu$ term in the denominator of Eq.16 cancels the $\lambda^3$ term of the integrated intensity.

However, the mass absorption coefficient, $\mu_m$, term in the exponent of Eq.16 must also be represented as by $\mu_m = c_\mu \lambda^3$ and the $\lambda$ replaced by Bragg's law that leads to another $\sin^3\theta$ factor in the exponent. Following these substitutions, the absorption factor for corrections of broad peaks is given by

$$A_{\alpha 2\theta}(\lambda) = \left( \frac{1 - e^{-8\, c_\mu\, \rho\, d_{hkl}^3\, \sin^3\theta\, k_\alpha t}}{c_\mu\, \rho\, k_\alpha\, \sin\alpha} \right) \qquad (17)$$

It is readily seen that the absorption factor for integrated intensity, Eq.16, and for peak broadening corrections, Eq.17, are different in thin film GIXRD, just as in symmetric diffraction. Therefore, the current use of Eq.16 for absorption corrections of higher angle broad peaks is incorrect. Eq.17 must be used for this purpose.

Eq.17 implies that for absorption corrections of higher angle peaks, d$_{hkl}$ must be used. This is similar to the case for symmetric diffraction and the same difficulties discussed earlier will be encountered.

The second aspect is the determination of the linear regime. In GIXRD, in the linear regime, the absorption factor, after linear approximation of Eq.16, is given by $A^{lr}_{\alpha 2\theta} = t/sin\alpha$. Hence, the absorption factor is independent of the absorption coefficient and also does not contribute to angle dependence.

However, Eq.17 must be used for absorption correction of broad peaks. After linear approximation of Eq.17 and combining with the $LP_{\alpha 2\theta}$ factor, Eq.15, the LPA factor in the linear regime is given by

$$LP^{lr}_{\alpha 2\theta} = sin\theta\,(1 + cos^2 2\theta) \tag{18}$$

The same result can also be obtained by starting with the LP and absorption factors for monochromatic radiation. We note that the absorption factor for monochromatic radiation is independent of the absorption coefficient in the linear regime. Hence, the only factors to be considered for LPA corrections are the LP factor and the $\lambda^3$ dependence of the integrated intensity that must be substituted with Bragg's law leading to $sin^3\theta$ dependence. Combining these two angular factors leads to the angular factor given by Eq.18.

The thickness range of validity of the linear regime is given by $\tau_{lr} = 0.135/\mu k_\alpha$. It extends to $\tau_{lr} =$ 2 nm - 200 nm ($\leq$ 1% error) depending on the material and the grazing angle for Cu-K$_\alpha$ radiation. The range can be extended by changing the radiation, as discussed above. Clearly, the linear regime is of practical importance in GIXRD as well.

**9.0 Discussion**

The higher angle peaks are broadened due to the presence of the Kα doublet in laboratory diffractometers. In this article we have extended the results for angular factor corrections of broad peaks obtained by Pike and Ladell (Pike, 1959; Ladell, 1961; Pike and Ladell, 1961) for the case of bulk materials to thin films.

From the extension of the results of Pike and Ladell to thin films (Pike, 1959; Ladell, 1961; Pike and Ladell, 1961) it follows that, to correct for broad peaks due to the presence of spectral profile in the Kα doublet, the main factors to be considered for LPA corrections are i) the λ dependence of the integrated intensity that must be substituted with Bragg's law ii) the LP factor and iii) the λ dependence of the mass absorption coefficient. In addition, any un-cancelled λ term must be substituted with Bragg's law since λ is not constant (Pike, 1959).

These ideas are general and are valid for several thin film diffraction techniques. We have discussed the absorption factors for symmetric and asymmetric diffraction. Other thin film diffraction techniques (Birkholz, 2005; Dummer *et al*. 2000; Liu et al. 2015; Wronski *et al*. 2009) have different expressions for the absorption factor. In these cases as well, the absorption factor has two terms for the linear absorption coefficient, μ, for which the above procedure must be applied. It follows that in these cases as well, the currently used absorption factor in peak profile studies is incorrect and the correct absorption factor obtained following the above procedure must be used.

The second result that follows from the analysis of the absorption factor is the presence of the linear regime for very thin films. This is also general and applicable to several thin film diffraction techniques (Birkholz, 2007; Vaudin, 1998; Vaudin, 1999). The thickness range of validity can be determined appropriately. It is of practical advantage, especially in LPA corrections of broad peaks, in that it eliminates the determination of several thin film parameters (thickness t, density ρ and interplanar spacing $d_{hkl}$) that are necessary in the thin film regime.

Thus, the two main results of the present work are of general applicability to many thin film diffraction techniques.

## 10.0 Conclusion

The angular (Lorentz, polarization and absorption (LPA)) corrections for higher angle broad peaks, due to the Kα doublet, in peak profile studies of bulk materials are based on the results of Pike and Ladell (Pike, 1959; Ladell, 1961; Pike and Ladell, 1961). These results have been extended to thin film x-ray diffraction. The current practice of using the absorption factors for monochromatic radiation in LPA corrections is incorrect. The correct absorption factors to be used for LPA corrections in symmetric and grazing incidence x-ray diffraction are derived. For very thin films, a linear approximation to the absorption factor can be made. In the linear regime, the absorption coefficient and the integrated intensity are independent of the absorption coefficient. Thus, based on thickness, materials are classified into three categories, bulk, thin film and linear regimes. The expressions for the linear regime are practically advantageous over the expressions in the thin film regime. Both the main results obtained from new analysis of the thin film absorption factors are general and can be extended to several thin film diffraction techniques.

**References**

Birkholz, M. (2005). *Thin Film Analysis by X-Ray Scattering*. Weinheim: Wiley-VCH.

Birkholz, M. (2007). *J. Appl. Cryst.* **40**, 735–742.

Breiby, D. W., Bunk, O., Andreasen, J. W., Lemke, H. T. and Nielsen, M. M. (2008). J. Appl. Cryst. **41**, 262–271

Creagh, D. C. and Hubbell, J. H. (2006). *International Tables for Crystallography,* Vol. C, Table 4.2.4.3, pp. 230-23.

Cullity, B. D. (1978) *Elements of X-ray Diffraction*, Reading: Addison Wesley, 2$^{nd}$ Ed.

Delhez, R., Mittemeijer, E. J., de Keijser, T. H. and Rozendaal, H. C. F. (1977). *J. Phys. E: Sci. Instrum*., **10**, 784-785.

Dümmer, T., Eigenmann, B. and Löhe, D. (2000). *Mater. Sci. Forum*, **321–324**, 81–86.

Ely, T., Predecki, P.K., and Noyan, I.C. (1999). *Adv. X-Ray Anal*. **41**, 467-478

Fewster, P. F. (1996). *Rep. Prog. Phys.*, **59** 1339-1407.

Fitzpatrick, M. E., Fry, A. T., Holdway, P., Kandil, F. A., Shackleton, J. and Suominen, L. (2005). *Measurement Good Practice Guide* No. 52, National Physical Laboratory, Teddington, UK, p. 64

Gloaguen, D., Fajoui, J. and Girault, B. (2014). Acta Met., **71** 136-144

Guazzone, F., Payzant, E.A., Speakman, S.A., and Ma. Y.H., (2006) *Ind. Eng. Chem. Res.*, **45**, 8145–8153.

Guinebretiere, R. (2007). *X-Ray Diffraction by Polycrystalline Materials*. London: ISTE Publishing.

Iwashita, N., Park, C. R., Fujimoto, H., Shiraishi, M. and Inagaki, M. (2004). *Carbon* **42** 701-714.

Ladell, J. (1961). *Acta Cryst.,* **14**, 47-53.

Lhotka, J., Kuzel, R., Cappuccio, G. and Valvoda, V. (2001). *Surf. Coat. Technol.*, **148,** 96-101.

Liu, Y, Bhamji, I., Withers, P.J., Wolfe, D.E., Motta, A.T. and Preuss, M. (2015). *J. Nucl. Mater.* **466**, 718–727.

Maret, M., Brombacher, C., Matthes, P., Makarov, D., Boudet, N. and Albrecht, M. (2012). *Phys. Rev. B* **86**, 024204.

McCusker, L. B,. Von Dreele, R. B., Cox, D. E, Louer, D. and Scardi, P. (1999). *J. Appl. Cryst.*, **32**, 36-50.

Miyazaki, H., Suzuki, T., Yano, T. and Iseki, T. (1992). *J. Nuclear Sci. Tech.*, **29**, 656-663.

Pecharsky, V. K. and Zavalij, P. Y. (2009). *Fundamentals of Powder Diffraction and Structural Characterization of Materials*, New York: Springer, 2$^{nd}$ Ed.

Pike, E. R. (1959). *Acta Cryst.,* **12**, 87-92.

Pike, E. R. and Ladell, J. (1961). Acta Cryst., **14**, 53-54.

Qiao, Z., Latz, R. and Mergel, D. (2004). *Thin Solid Films,* **466**, 250-258.

Skrzypek, S. J. and Baczmański, A. (2001). *Adv. X-Ray Anal*. **44**, 134-145

Somers, M. A. J., van der Pers, N. M., Schalkoord, D., and Mittemeijer, E. J. (1989). *Met. Trans. A*, **20A**, 1533-1539.

Tian, H. H. and Atzmon, M. (1999). *Phil. Mag. A*, **79**, 1769-1786

Tian, X. and Zhang, Y. (2009). *Mater. Sci. Eng.* A, **516**, 73-77.

Vaudin, M. D., Rupich, M. W., Jowett, M., Riley, G. N. and Bingert, J. F. (1998). *J. Mater. Res*. **13**, 2910-2919.

Vaudin, M. D. (1999) in Proceedings of the 12th International Conference on Textures of Materials, ICOTOM-12, pp. 186-191.

Volz, H. M., Vogel, S. C., Necker, C. T., Roberts, J. A., Lawson, A. C., Williams, D. J., Daemen, L. L., Lutterotti, L. and Pehl, J. (2006). *Powder Diffr.*, **21**, 114-117

Warren, B. E. (1990) *X-ray Diffraction.* New York: Dover, 2nd Ed.

Waseda, Y, Matsubara, E. and Shinoda, K. (2011). *X-Ray Diffraction Crystallography.* Heidelberg: Springer.

Wroñski, S., Wierzbanowski, K., Baczmañski, A., Lodini, A., Braham, C. and Seiler, W. (2009). *Powder Diffr. Suppl.*, **24**, S11-S15

Yang, E., Laughlin, D. E. and Zhu, J. G. (2012). *IEEE Trans. Magn*. **48**, 7-12.

# List of Tables

**Table I** Summary of Lorentz-Polarization and absorption factors for bulk, thin film and linear regimes for monochromatic radiation and for angular corrections for Kα doublet.

**Table II** Primary data for calculating the relative intensities in FePt thin films (from Yang *et al.,* 2012). Additional quantities calculated from primary data using present approach.

# List of Figures

**Fig.1**. Comparison of the thin film absorption factor (1-exp(-2μt/sinθ)) (red) and its linear approximation (2μt/sinθ) (blue) for ZnO(100). The value of the mass absorption coefficient used is 1.01 $\mu_m(\bar{\alpha})$. The linear regime is upto 680 nm.

**Table I**

|  | **LPA factor** (integrated intensity, monochromatic radiation) | **LPA factor** (angular corrections for Kα doublet) |
|---|---|---|
| **Bulk regime** | $\left(\dfrac{1+\cos^2 2\theta_{hkl}}{\sin^2\theta_{hkl}\,\cos\theta_{hkl}}\right)$ | $\left(\dfrac{1+\cos^2 2\theta_{hkl}}{\sin^2\theta_{hkl}\,\cos\theta_{hkl}}\right)$ |
| **Thin film regime** | $\left(\dfrac{1+\cos^2 2\theta_{hkl}}{\sin^2\theta_{hkl}\,\cos\theta_{hkl}}\right)$ $\left(1-e^{-\dfrac{2\mu_m(\bar{\alpha})\rho t}{\sin\theta_{hkl}}}\right)$ | $\left(\dfrac{1+\cos^2 2\theta_{hkl}}{\sin^2\theta_{hkl}\,\cos\theta_{hkl}}\right)$ $(1-e^{-16\,c_\mu\,\rho\,d_{hkl}^3\,\sin^2\theta\,t})$ |
| **Linear regime** | $\left(\dfrac{1+\cos^2 2\theta_{hkl}}{\sin^3\theta_{hkl}\,\cos\theta_{hkl}}\right)$ | $\left(\dfrac{1+\cos^2 2\theta_{hkl}}{\cos\theta_{hkl}}\right)$ |

**Table II**

| hkl | (001) | (002) |
|---|---|---|
| 2θ | 23.99 | 49.09 |
| sinθ/λ | 0.135 | 0.269 |
| $\|F\|^2$ | 9659 | 24665 |
| Temp Factor | 0.9643 | 0.8649 |
| Abs factor (for thin films) = (1-exp-2μt/sinθ) | $1 - e^{-0.0032t}$ | $1 - e^{-0.0016t}$ |
| $LP_{hkl}$ (Eq.2) | 43.435 | 9.102 |
| $LP_{hkl}^{lr}$ (Eq.11) | 208.99 | 21.91 |
| multiplicity | 2 | 2 |
| $I_{hkl}$ (Eq.1) | 809122 | 388342 |
| $I_{hkl}^{lr}$ (Eq.10) | 3893138 | 934801 |
| $\dfrac{I_{hkl}(001)}{I_{hkl}(002)}$ | 2.08 ||
| $\dfrac{I_{hkl}^{lr}(001)}{I_{hkl}^{lr}(002)}$ | 4.16 ||

**Figure 1**

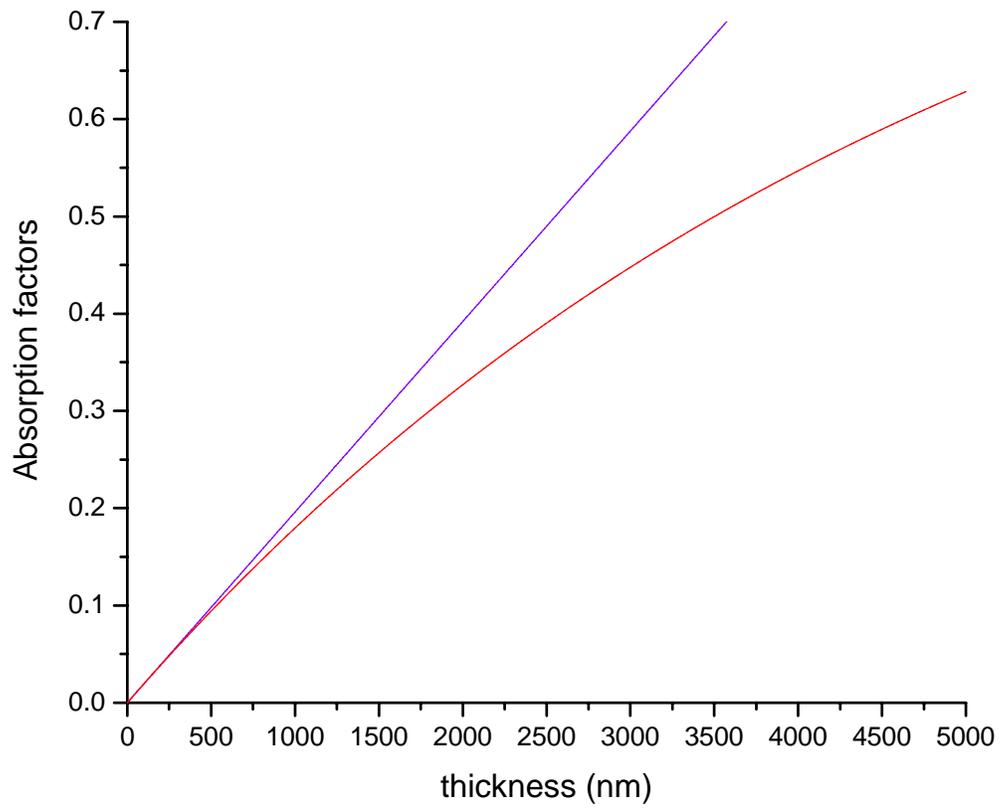